\documentclass[twocolumn]{aastex631}
\usepackage{amsmath}
\usepackage{xspace}

\usepackage{multirow}

\newcommand{\cxo}{{\it Chandra}}
\newcommand{\xmm}{{\it XMM-Newton}\xspace}

\newcommand{\revI}[1]{{#1}}

\begin{document}


\title{\revI{Detection of Anisotropies in the Circumgalactic Medium of Disk Galaxies: \\ Supermassive Black Hole Activity or Star Formation-driven Outflows?}}

\correspondingauthor{andrea.sacchi@cfa.harvard.edu}
\author[0000-0002-7295-5661]{Andrea Sacchi}
\author[0000-0003-0573-7733]{\'Akos Bogd\'an}
\affiliation{Center for Astrophysics $\vert$ Harvard \& Smithsonian, 60 Garden Street, Cambridge, MA 20138, USA}
\author[0000-0003-4983-0462]{Nhut Truong}
\affiliation{NASA Goddard Space Flight Center, Greenbelt, MD 20771, USA}
\affiliation{Center for Space Sciences and Technology, University of Maryland, 1000 Hilltop Circle, Baltimore, MD 21250, USA}







\begin{abstract}
Gamma and X-ray observatories have revealed spectacular structures in the emission of the tenuous hot gas surrounding the Milky Way (MW), known as the Fermi and eROSITA bubbles. Galaxy formation simulations suggest that MW-like bubbles could be ubiquitous, but their emission may be too faint to detect with today’s instruments in individual external galaxies. In this paper, we present an analysis of stacked \textit{Chandra} observations of 93 nearby galaxies.
We detected soft, diffuse X-rays from the CGM, extending up to 14~kpc, with a luminosity of $(4.2\pm0.7)\times10^{39}$~erg/s in the $0.3-2$~keV band. To probe its spatial distribution, we constructed an azimuthal profile and found a significant enhancement along the galactic minor axis. When dividing our sample by stellar mass, central supermassive black hole mass, and star formation rate, we found that only high star formation rate galaxies exhibit significant anisotropies in the CGM emission.
To investigate whether the observed anisotropies could be attributed to MW-like bubbles, we compared our results with TNG50 simulations. In these simulations,  X-ray bubbles are strongly correlated with mass of the central supermassive black hole and typically extend to much larger, $\sim50$~kpc, scales.
We conclude that the observed anisotropies are either caused by AGN-driven MW-like bubbles confined to smaller, $\sim10$~kpc, scales, or by star formation- or starburst-driven bubbles/outflows.  
\end{abstract}

\keywords{galaxy nuclei -- circumgalactic medium -- high energy astrophysics -- X-ray astronomy}

\section{Introduction} \label{sec:intro}

Most galaxies host diffuse gaseous material that is not part of their disk and does not compose the interstellar medium. This material is called circumgalactic medium (CGM). The current understanding of galaxy-formation processes suggests that the CGM is composed of intergalactic material that accumulates in dark matter halos. These same dark matter halos host at their centers the systems of stars and gas we call galaxies \citep{white78b,white91}.
While the bulk of the CGM remained in the extended dark matter halos, when galaxies started forming, part of the gas cooled down and sank in the galactic disc, providing material to fuel star formation and thus building up the stellar component of galaxies. This means that studying the CGM offers unique insights into the physical processes that determine the appearance and properties of galaxies, supermassive black holes (SMBHs) and stellar feedback, metal enrichment processes, and baryon and metal recycling \citep[see][for reviews]{tumlinson17,werner19}. As stated, the majority of the gas stays in the dark matter halo. For Milky Way-mass and more massive galaxies, the bulk of the CGM is in the hot, X-ray-emitting, phase, shock-heated up to millions of Kelvin. This high-energy domain is hence a privileged window into the large-scale properties of the CGM.

\begin{figure}[t!]
\centering
\includegraphics[width = 0.5\textwidth]{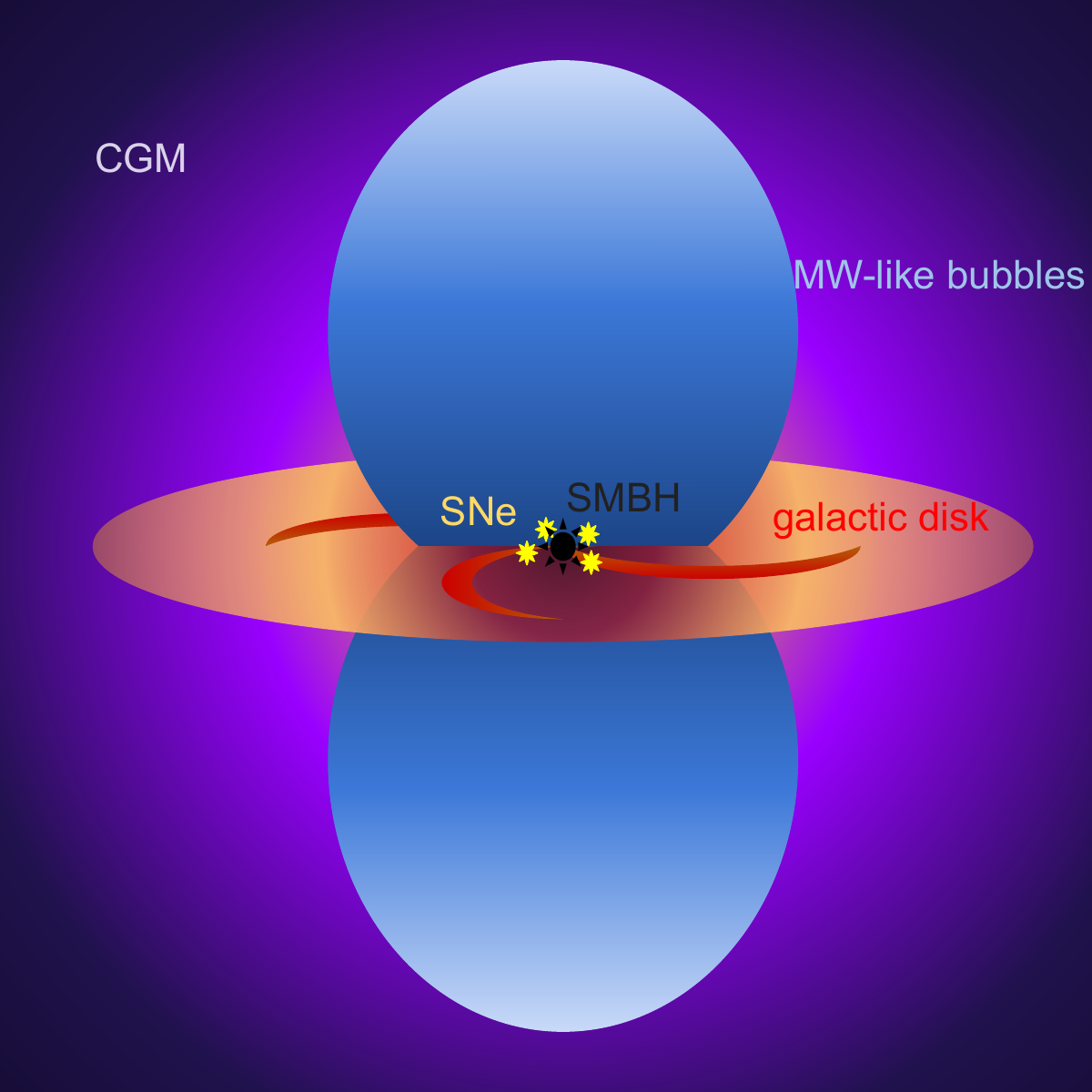}
\caption{A schematic representation of MW-like bubbles (blue) rising above the galactic disk (yellow-red) in an approximately edge-on view. The system is embedded in a spherically symmetric large-scale CGM (purple to violet). The SMBH \revI{and SNe}, in the galactic center are also indicated. \label{fig:bubs}}
\end{figure}

More than a decade ago, the Fermi Gamma-ray Space Telescope uncovered large gamma-ray emitting regions that extend to $\sim$10~kpc height above and below the Galactic plane, rising from the Galactic center \citep{su10}. Recently, the all-sky survey performed by eROSITA (extended ROentgen Survey with an Imaging Telescope Array, \citealt{predehl21}, the main instrument aboard the German-Russian satellite \textit{Spektrum Roentgen Gamma} \citealt{sunyaev21}) identified shell-like structures in the soft X-ray band, which have a spatial structure similar to the Fermi bubbles \citep{predehl20}, albeit somewhat more extended. This X-ray feature, the eROSITA bubbles, stretches $\sim$14~kpc both above and below the Galactic plane (see Fig.\ \ref{fig:bubs} for a schematic view) and exhibits an X-ray luminosity of $\approx10^{39}$~erg/s. The origin of the eROSITA/Fermi bubbles, hereafter Milky Way (MW) bubbles, is debated: they may either be inflated by the energetic feedback from Milky Way's SMBH, or they may originate from the energetic winds from intense star formation \citep[e.g.][]{sarkar15,yang22,sarkar23,gupta23,li.chegnzhe24,sarkar24}.

The detection of MW bubbles raises a critical question: do other galaxies also host MW-like bubbles? Modern galaxy formation simulations suggest that MW-like bubbles and/or anisotropies should be common in disk galaxies. For example, a detailed analysis of the TNG50 simulation revealed that a substantial fraction of the Milky Way mass and more massive galaxies should host MW-like bubbles \citep{pillepich21}. In the simulation, these bubbles originate from the kinetic energy injection of SMBHs that accrete at low Eddington ratios. Similarly, the EAGLE simulation also predicts anisotropies, though these are less pronounced and are observable at larger galactocentric distances ($\sim40-60$~kpc) \citep{truong21,nica22}. So while simulations agree that MW-like bubbles or anisotropies should exist around external disk galaxies, their presence has yet to be observationally proven.

Some anisotropies have been detected in the emission of the CGM surrounding lower-mass disk galaxies, like M31 \citep{bogdan08} or M104 \citep{li11}. These outflows originate from the galactic bulges and extend over a few kpc scales (much smaller than the observed MW bubbles), and are due to type Ia supernovae. Bubble-like structures and X-ray cavities have also been detected around massive elliptical galaxies \citep[e.g.][]{bogdan14,randall15,ubertosi25}. In these cases, such features are associated with AGN-driven outflows and jets, as highlighted by the presence of radio lobes that fill the X-ray cavities.

Thus far, however, \cxo\ and \xmm\ have only detected the large-scale CGM around a handful of individual massive disk galaxies \citep{anderson11,bogdan13,anderson15,bogdan17,lijt17}, without evident signatures of asymmetries in the distribution of their CGM. Another approach to detecting the faint and diffuse emission from CGM, enabled by large-area X-ray surveys, is stacking images of multiple galaxies. This exercise has been performed exploiting ROSAT data \citep{anderson15}, and more recently on eFEDS data \citep{comparat22,chadayammuri22}, and eROSITA data \citep{zhang24a}. These efforts successfully detected the large-scale CGM, but the stacking procedure averaged any asymmetry in the CGM spatial distribution, preventing the detection of any anisotropy or the presence of MW-like bubbles. Furthermore, the typical redshift of the galaxies in said samples ($z\approx0.1$) and the large point spread function of eROSITA ($\approx30"$) prevents resolving anisotropies on the typical scales of the MW bubbles.

\begin{figure*}[t]
\centering
\includegraphics[width=0.70 \textwidth]{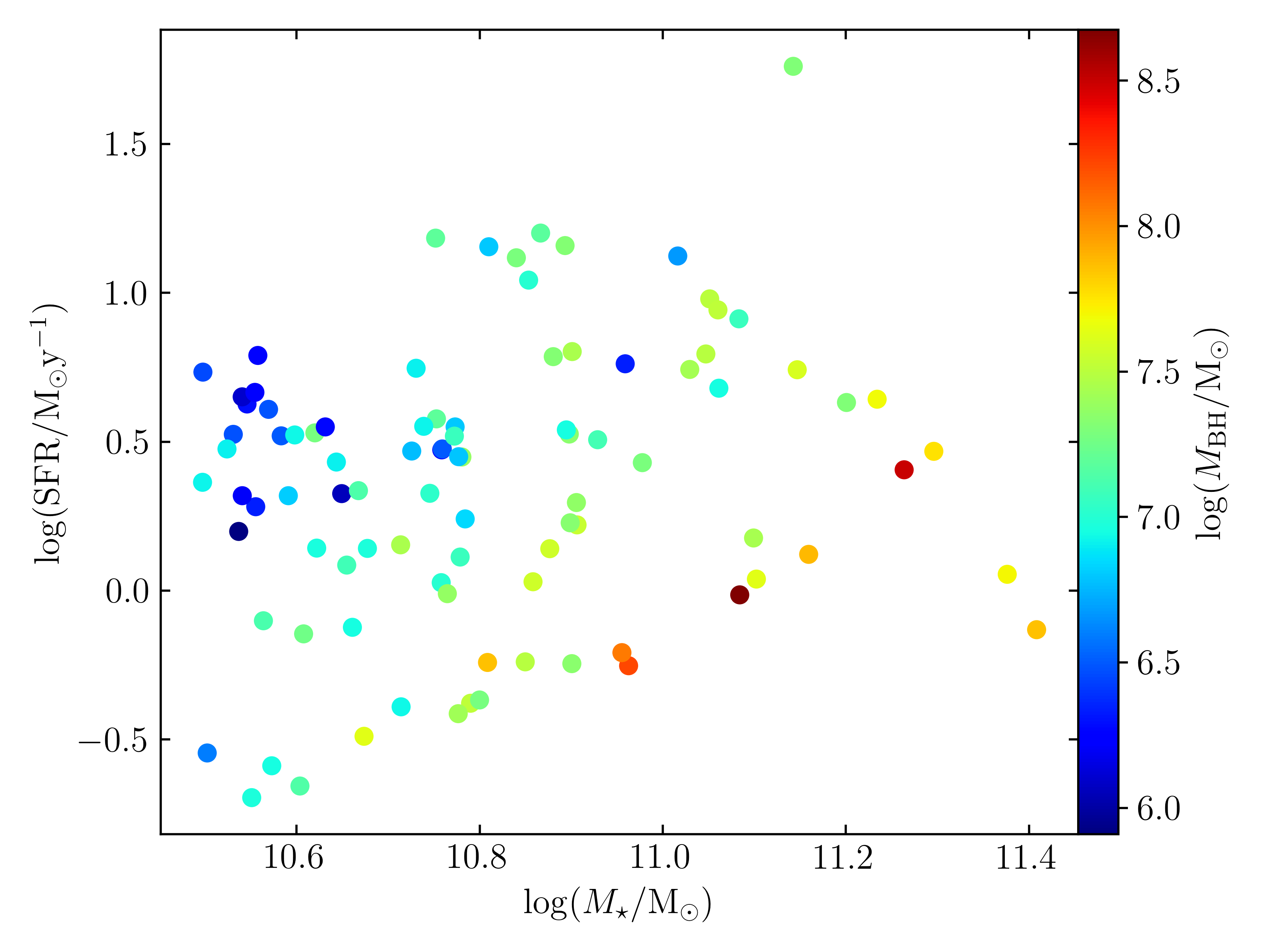}
\caption{Relationship between the star formation rate and the 
stellar mass for the 93 galaxies in our sample, as obtained from the HECATE catalog. The colorbar indicates the inferred mass of the SMBH hosted in each galaxy.}
\label{fig:sam_prop}
\end{figure*}

Detecting and mapping the CGM around MW-mass disk galaxies is challenging because of the low density and temperature of their CGM. Given their typical surface brightness, detecting the inner regions ($<30$~kpc) of the CGM would require $\gtrsim10^6$ s exposure time with present-day observatories (e.g.\ \cxo, \xmm\, or eROSITA) \citep{oppenheimer20,pillepich21,truong21}. Thus, the presence and ubiquity of MW-like bubbles around a substantial sample of MW-mass disk galaxies remains an open question.

To probe the existence of MW-like bubbles with publicly available data, we leverage archival \textit{Chandra} observations of 93 edge-on disk galaxies with a total exposure time of $\approx2.2$~Ms. By stacking the X-ray photons from these systems, exploiting \cxo\ sensitivity and unmatched angular resolution, we were able to not only detect the CGM but also to probe its spatial structure thereby revealing the presence of anisotropies in the CGM emission. We explored the correlation of these features with the star formation rates (SFR), stellar mass, and SMBH mass of the galaxies. We compare our results with the TNG50 simulations to investigate the possible origins of these anisotropies.

The paper is organized as follows: in Section \ref{sec:sample} we describe how we built the sample and cleaned the X-ray images of the disk galaxies we employed in our analysis. In Section \ref{sec:bubs} we describe how we analyzed our sample to detect the CGM emission and MW-like bubbles. In Section \ref{sec:sim} we show our comparison with mock X-ray images of simulated galaxies. In Section \ref{sec:disc} we discuss our results and the possible origins of the observed anisotropies in the CGM emission, and in Section \ref{sec:conc} we draw our conclusions.

\section{The sample}
\label{sec:sample}

\subsection{Sample selection}

To detect the CGM and probe the existence and ubiquity of MW-like bubbles/anisotropies, a well-defined galaxy sample must be defined. Our parent sample is drawn from the HECATE catalog \citep{kovlakas21}, which is an all-sky catalog containing over 200,000 local galaxies within $D<200$~Mpc ($z<0.047$). From this, we selected edge-on ($i>75^\circ$) disk galaxies with a stellar mass of $\log (M_{\rm \star}/\rm{M_\odot}) > 10.5$ and measured SFR.

We chose to limit our selection to edge-on disk galaxies to probe the structure of the CGM above and below the galactic disk avoiding projection effects. We set a lower limit on the galactic stellar mass to ensure that the galaxies in our sample have a sufficiently deep potential well to retain hot, X-ray-emitting gas. No upper limit was imposed on stellar mass, allowing us to explore the presence of MW-like bubbles across the entire mass range. Note that the TNG50 simulations predict that more massive galaxies are more likely to host MW-like bubbles. At this stage, 8321 galaxies satisfy these criteria.

We cross-matched this galaxy sample with the \textit{Chandra} archive and identified 581 galaxies with publicly available X-ray data. From this pool, by visual inspection, we excluded 147 galaxies that were too close to the edge of the detector ($\lesssim35$~kpc of projected distance), which would prevent the full imaging of MW-like bubbles or anisotropies, and the extraction of a local background from its surrounding regions. We also excluded, through visual inspection, 341 galaxies that were members of, or projected onto, galaxy groups or clusters. The rich environments of galaxy groups and clusters are characterized by bright X-ray emission that would outshine the CGM of individual galaxies and bias the results of the stacking analysis. After excluding these galaxies, our final sample included 93 disk galaxies, with a total of 133 individual observations amounting to 2,245~ks of exposure time. The average exposure time of a single observation is 18~ks, the longest exposure amounts to 118~ks, while the shortest is 1.5~ks. The physical properties of the galaxies in our sample, including ObsIDs and exposure times are reported in Table~\ref{tab:prop}.

\begin{figure}[t]
\centering
\includegraphics[width = 0.5\textwidth]{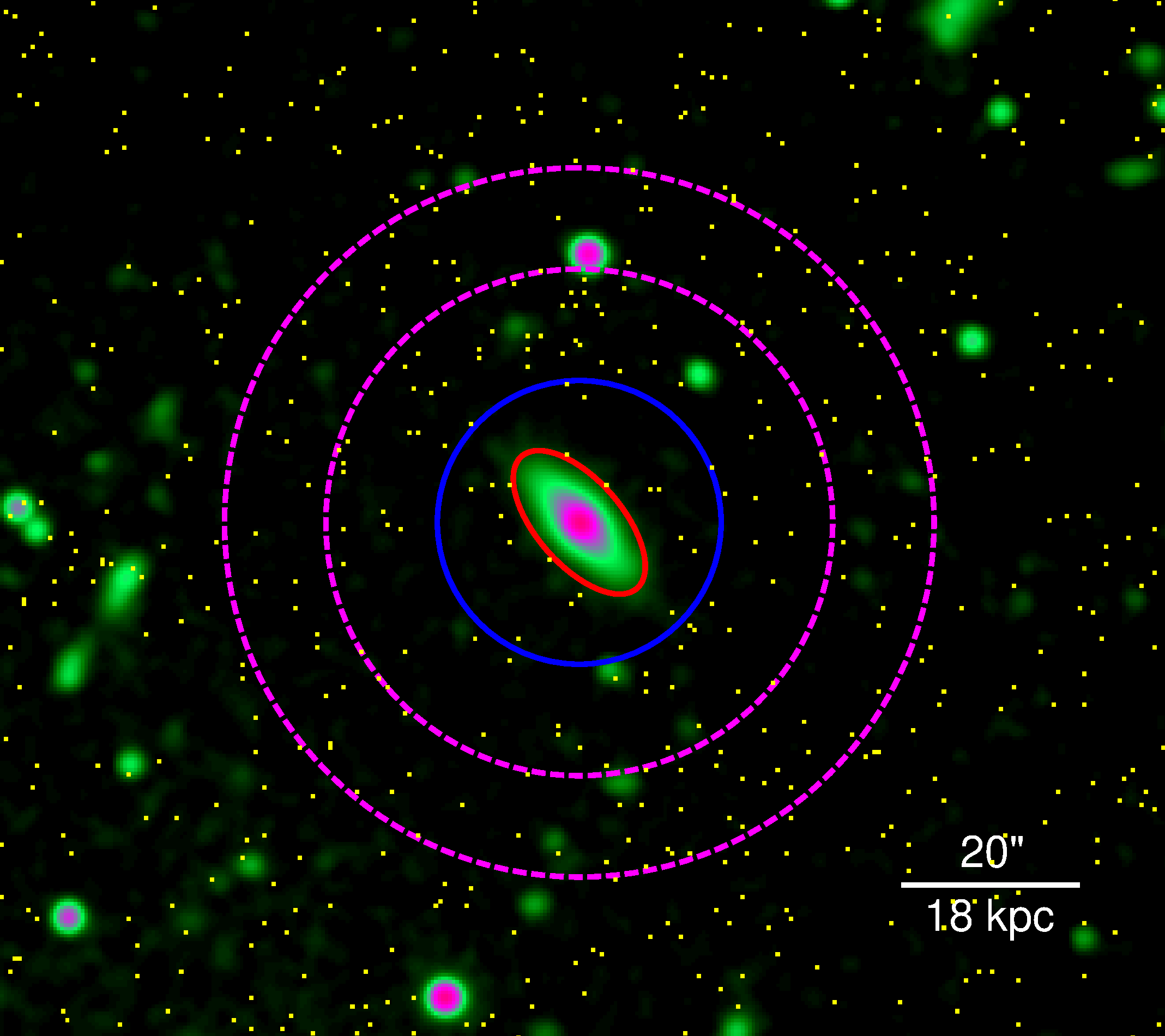}
\caption{Composite optical (SDSS $r$-band) and X-ray (Chandra) image of a representative galaxy, GAMA~79711, from our sample. The red ellipse marks the excluded region, which contains $\gtrsim95\%$ of the stellar light. Source and background counts were extracted from the blue circle ($14$~kpc radius) and magenta annulus ($25-35$~kpc radii), respectively. Yellow pixels indicate individual X-ray counts detected by \textit{Chandra}.
 \label{fig:scheme}}
\end{figure}

The bulk of our sample ($\approx60\%$) is composed of MW-mass galaxies with $M_{\rm \star} = (3-7)\times 10^{10} \ \rm{M_{\odot}}$, while the ten most massive galaxies have stellar masses in the range of $M_{\rm \star} = (1-3)\times 10^{11} \ \rm{M_{\odot}}$. The average distance of the galaxies is $\approx130$ Mpc (at this distance 15~kpc corresponds to a projected distance of $\approx25"$) and their star formation rate (SFR) spans about two orders of magnitude, from $0.3-50 \ \rm{M_{\odot} \ yr^{-1}}$ (see Fig.\ \ref{fig:sam_prop}). Finally, we infer the mass of the central SMBHs by adopting the $M_\textup{BH}-\sigma$ relation from \citet{mcconnell11}, where $M_\textup{BH}$ is the SMBH mass and $\sigma$ is the central stellar velocity dispersion obtained from the HyperLeda database \citep{makarov14}. For those galaxies for which no central velocity dispersion information was available, we computed the SMBH mass adopting the $M_\textup{BH}-M_\textup{gal}$ relation derived by \citet{reines15}. We adopted the relation for elliptical galaxies or AGN following the optical classification of each galaxy. The derived SMBH masses span from $\approx10^6$~M$_\odot$ to $\approx10^8$~M$_\odot$. Considering these ranges of galactic mass, SMBH mass, and star formation rate, the sample we built allows us to probe the presence of MW-like bubbles or anisotropies, as well as to investigate potential variations depending on one or more of these parameters.

\subsection{X-ray data analysis}\label{sec:proc}

For each galaxy in our sample, we retrieved all available observations from the \cxo\ archive\footnote{\url{https://cda.harvard.edu/chaser/mainEntry.do}} and reprocessed and reduced the data with the Chandra Interactive Analysis of Observations software package (\texttt{CIAO}, v.4.12; \citealt{fruscione06}) and the \texttt{CALDB} 4.9.0 release of the calibration files. 

For each observation, we generated point spread function images and weighted exposure maps. The exposure maps were weighted using a thermal plasma spectrum (\textsc{apec}), with a temperature of $kT=0.3$~keV and metal abundances of $Z=0.3 \ \rm{Z_{\odot}}$ \citep{wilms00}, which is the expected spectrum of the CGM given the dark matter halo mass of MW-like galaxies. This spectral profile was absorbed using a layer of Galactic absorption (\textsc{TBabs}) with a column density of $N_\textup{H}=2\times10^{20}$~cm$^{-2}$, the typical value for the galaxies in our sample.

From each X-ray image, we identified bright point sources with the \texttt{wavedetect} tool. To this end, we used the wavelet scales of 1.414, 2, 2.828, 4, 5.636, and 8, and the corresponding point spread function maps, and all other parameters were left as default. We proceeded to exclude all identified point sources, which could contaminate the diffuse emission.

To eliminate contamination from the population of unresolved X-ray binaries and the hot interstellar medium in the galactic disks, we masked each galaxy by removing an elliptical region. The major and minor axes of this region were determined using optical measurements from the HECATE catalog, which provides the D25 isophotal diameter. Based on the SDSS r-band images, we estimate that the fraction of stellar light outside the masked regions is, on average, $\lesssim5\%$ of the total optical emission. We assume the same fraction for the residual contamination from unresolved X-ray binaries. This is a conservative assumption as, while it is true that low-mass X-ray binaries are distributed as the stellar light, this is not true for high-mass X-ray binaries. These are concentrated in star-forming regions, usually clustered in the galactic plane, and hence a larger fraction of these will be removed via our masking procedure.

\section{Detecting CGM emission and anisotropies} \label{sec:bubs}

\begin{figure*}[!t]
\centering
\includegraphics[width=0.7\textwidth]{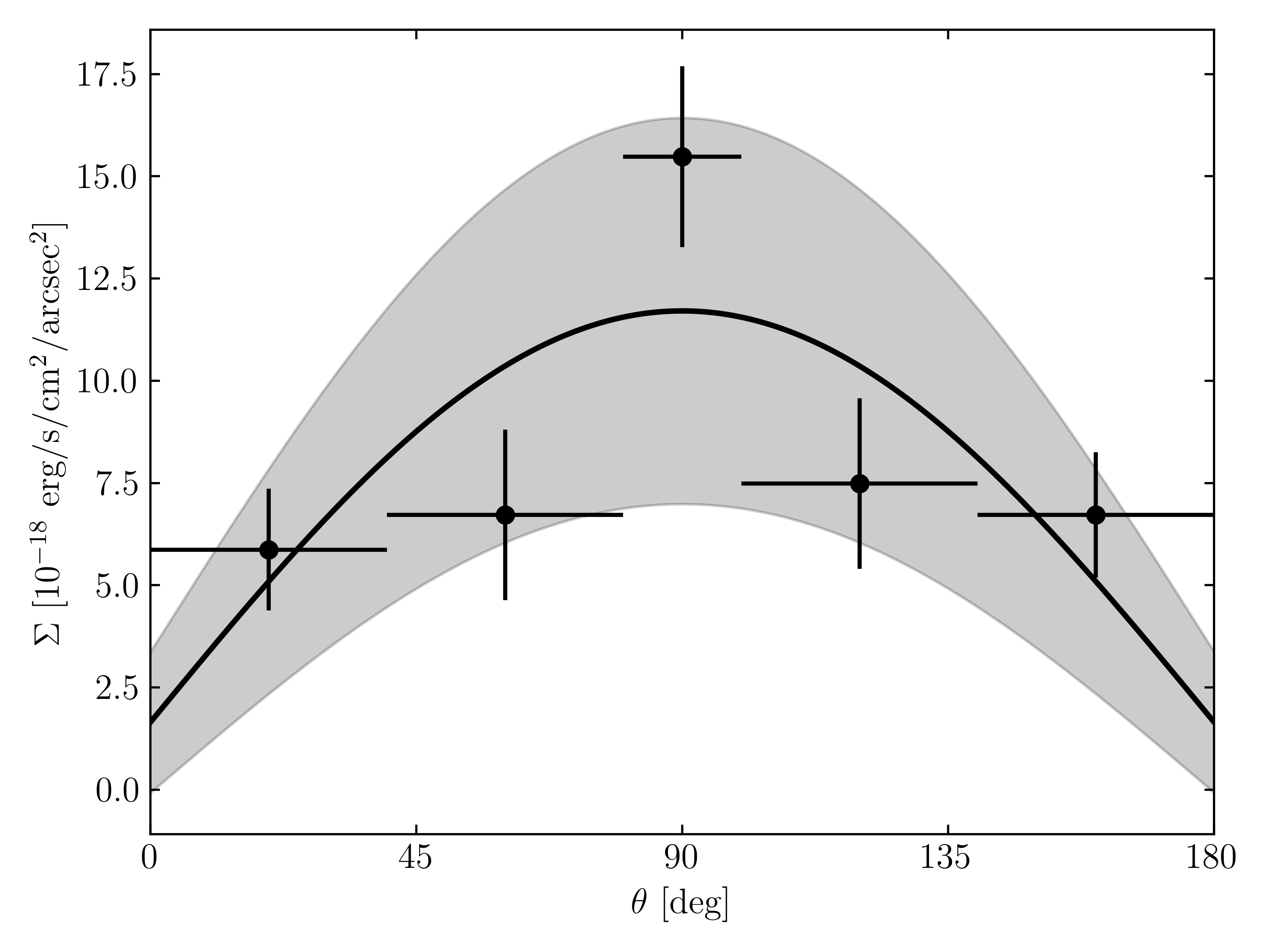}
\includegraphics[width=0.32 \textwidth]{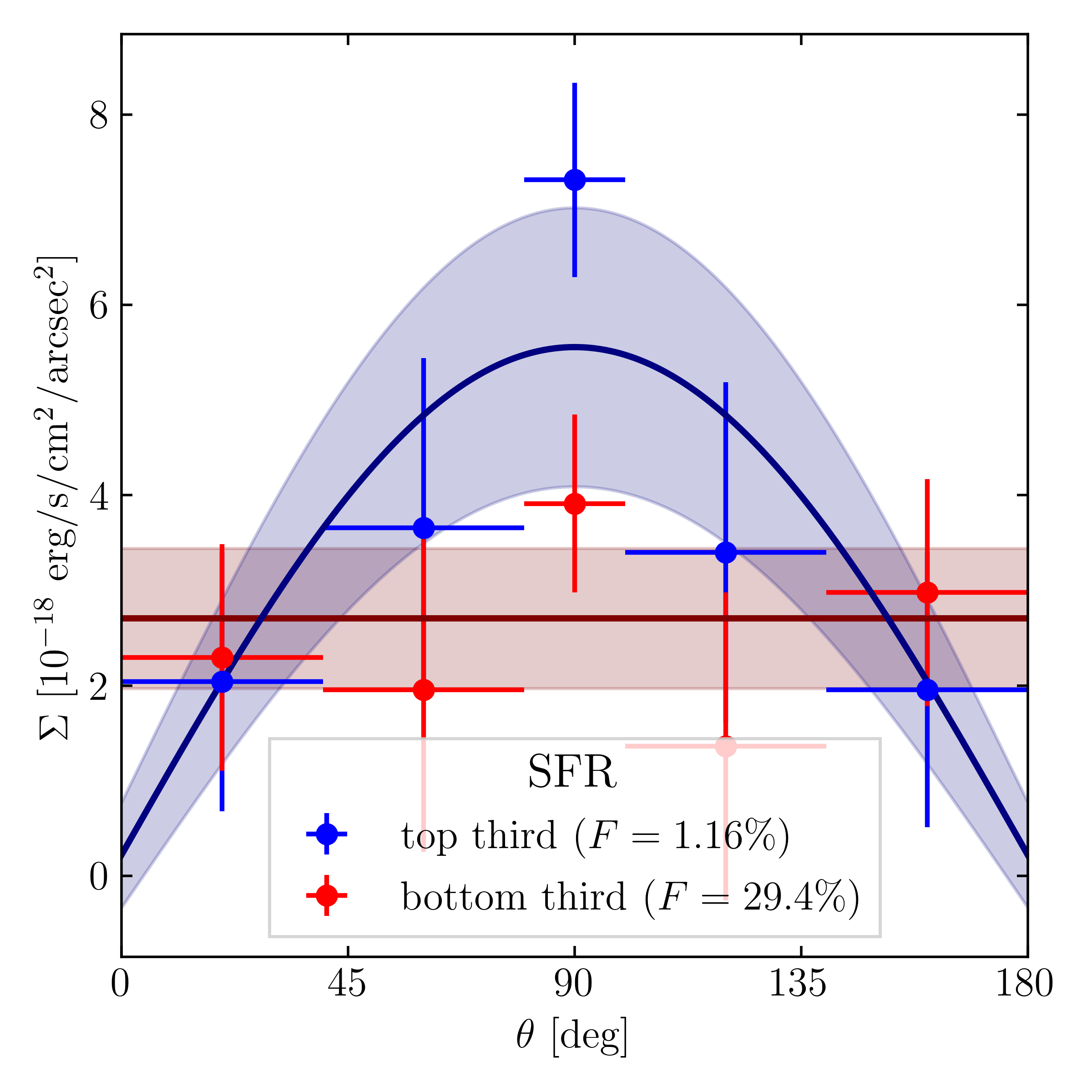}
\includegraphics[width=0.32 \textwidth]{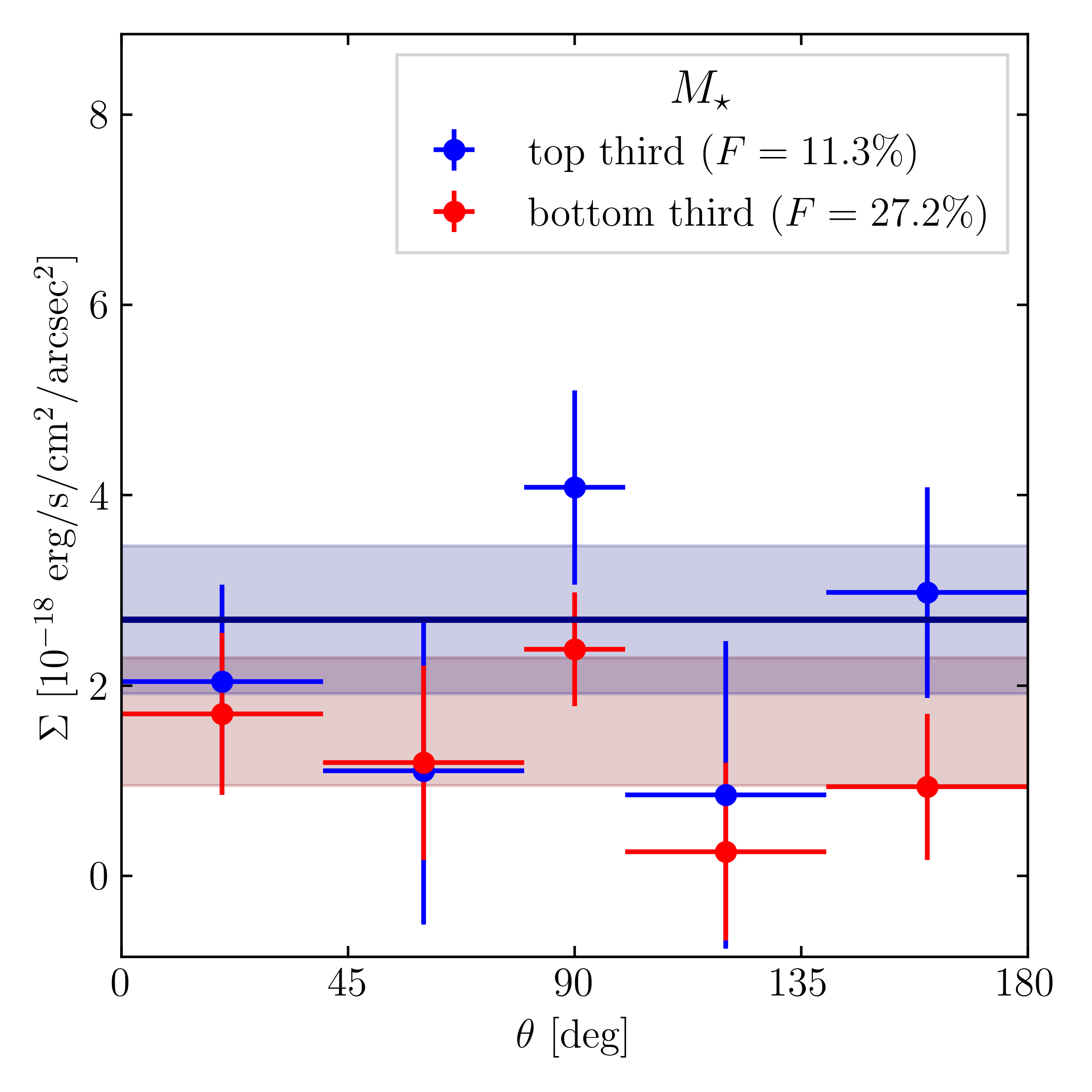}
\includegraphics[width=0.32 \textwidth]{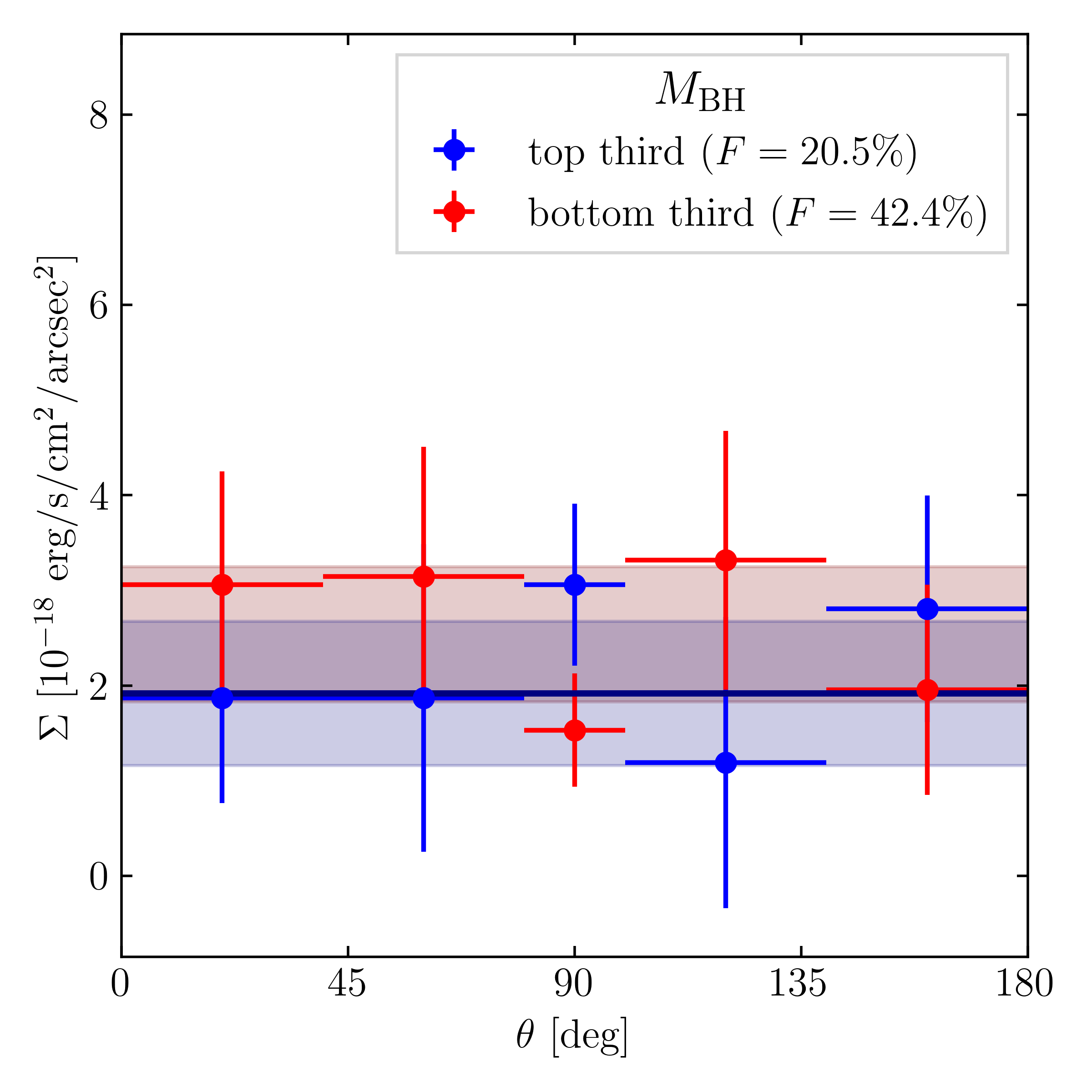}
\caption{Azimuthal profiles of the stacked X-ray images of the observed galaxies. The top panel shows the profile for the full sample. The second row presents the profiles for different subsamples, divided based on SFR (left panel), stellar mass (middle panel), and SMBH mass (right panel). Blue and red indicate the top and bottom third of each parameter, respectively. Solid lines represent the best fit, while the shaded regions show the $1\sigma$ uncertainties. \revI{In the legend, are also reported the results of the $F$-tests.}}
\label{fig:sig}
\end{figure*}

\subsection{Detecting the CGM}

To detect the CGM emission from each X-ray image, we extracted and co-added counts from a circular region centered on the galaxy centroid (the galaxy itself was masked as described above) with a radius of 14~kpc. This choice of radius corresponds to the dimensions of the MW bubbles and, at the same time, maximizes the S/N of the signal, as highlighted below. Background counts were extracted from annuli with $25-35$~kpc radii (see Fig.\ \ref{fig:scheme} for an example of this scheme). We extracted source and background X-ray counts in the soft (0.3--2~keV) and hard (3--8~keV) bands using \texttt{dmextract}, applying the corresponding exposure maps. These energy bands were chosen to ensure that the soft band captures thermal emission from gas at various temperatures across different galaxies in our sample, while the hard band excludes such thermal emission. 

None of the galaxies in the sample is individually detected in any considered band. Therefore, we stacked the X-ray photons from all the galaxies in our sample and detected a statistically significant signal in the $0.3-2$~keV band. We accumulated 2014 gross counts over 1625 background counts, corresponding to $389\pm60$ net counts, yielding a $\approx6.5\sigma$ detection \revI{of the CGM in the stacked data}. To calculate the luminosity of the detected emission, we first derived the average photon flux of the stacked sample using the exposure maps, which resulted in $(2.3\pm0.4)\times10^{-6}$~photons/s/cm$^2$. This photon flux was then converted into energy flux, adopting the same spectral profile used to weight the exposure maps, yielding an energy flux of $(2.2\pm0.4)\times10^{-15}$~erg/s/cm$^2$. Finally, using the mean distance of the galaxies, $\langle D\rangle=130$~Mpc, we calculated an average luminosity of $(4.2\pm0.7)\times10^{39}$~erg/s in the $0.3-2$~keV band.

We proceeded to extract the counts in the same regions, this time considering the $3-8$~keV X-ray band. We collected 2222 counts over a background of 2137, corresponding to $85\pm66$ net counts, which is consistent with a statistical fluctuation at the $\approx1.3\sigma$ level. The detection of the signal only in the soft band suggests the thermal origin of the emission. Indeed, if the emission detected in the soft band were due to unresolved high- and or low-mass X-ray binaries (resolved point-like sources were already excluded from the X-ray images as described above), their typical power-law spectra would produce a strong signal in the hard X-ray band as well. 
\begin{figure*}[t]
\centering
\includegraphics[width=0.7 \textwidth]{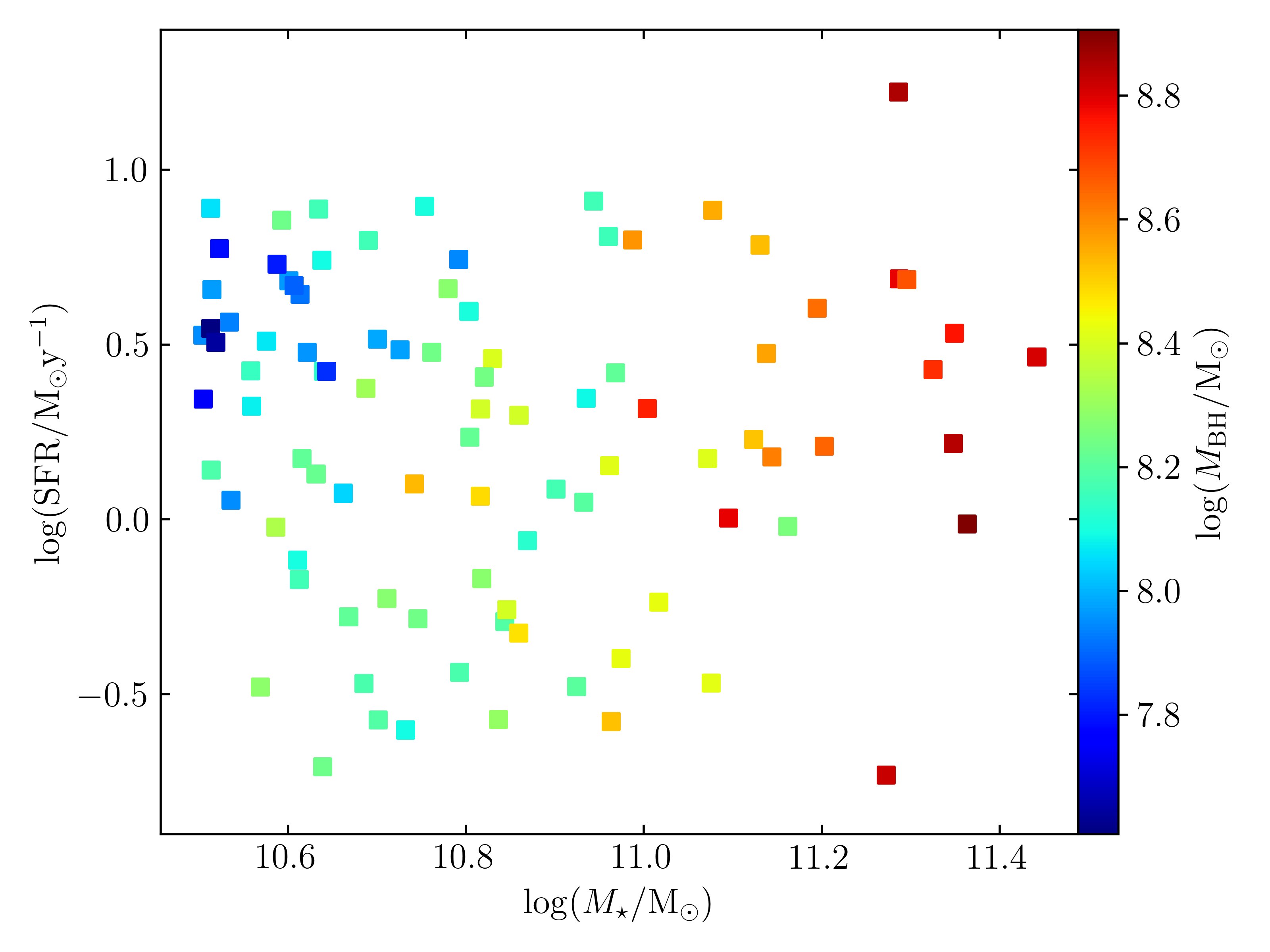}
\caption{Relationship between the star formation rate and stellar mass for the 93 galaxies simulated galaxy from TNG50. The color bar indicates the mass of the SMBH hosted in each simulated galaxy.}
\label{fig:mock_prop}
\end{figure*}

Unresolved populations of X-ray binaries would still contribute to some extent to the signal we detected in the soft band. To estimate this contribution, we extracted the counts and average luminosity in the hard $3-8$~keV band, but this time without masking out the galaxies. We obtained 240 net counts, corresponding to a luminosity of $1.8\times10^{39}$~erg/s. We converted this hard X-ray band emission into the expected luminosity in soft $0.3-2$~keV band by using the online tool \texttt{PIMMS}\footnote{\url{https://cxc.harvard.edu/toolkit/pimms.jsp}} assuming a power-law profile spectrum with photon index $\Gamma=2$, absorbed by a column density of $N_\textup{H}=2\times10^{20}$~cm$^{-2}$. In the soft $0.3-2$~keV band we expect a luminosity of $4.2\times10^{39}$~erg/s. Following the calculations described in Section \ref{sec:proc}, we consider that the residual emission, outside of the masked region, amounts to 5\% of the one computed, finally obtaining a luminosity of $2.1\times10^{38}$~erg/s. This value is a factor of 20 lower than the luminosity we obtained for the detected soft diffuse emission. This confirms that unresolved X-ray binaries do not significantly contaminate our observed X-ray signal in the $0.3-2$~keV band and that the signal we detected is due to the large-scale CGM surrounding the galaxies in our sample.

\subsection{Detecting CGM anisotropies}

By stacking the X-ray photons from the outskirts of the galaxies in our sample, we successfully detected extended soft X-ray emission originating from the large-scale CGM. Next, we explore possible anisotropies in the CGM spatial distribution.

To this end, we constructed the azimuthal profile of the CGM, centering it on the galactic centroids. To increase the signal-to-noise ratio, we combined counts extracted from ``above'' and ``below'' the galactic plane, where $\theta=0^\circ$ and $\theta=180^\circ$ correspond to the galactic plane, while $\theta=90^\circ$ indicates the region along the galactic minor axis. \revI{We selected five bins, one $20^\circ$ wide, centered on $\theta=90^\circ$, and four $40^\circ$ wide. Surface brightness was obtained for each bin using \texttt{dmextract} in the same fashion described above. This guarantees that the fact that we mask the galaxy with an elliptical region does not affect the shape of the azimuthal profile (although it might affect the statistical errors in the bins closer to the galactic plane).} The resulting azimuthal profile reveals an enhancement in the direction of the galactic minor axis ($\theta=90^\circ$), corresponding to the region above the galactic center. This profile, shown in the top panel of Fig.\ \ref{fig:sig}, is well reproduced by a sinusoidal model.

To explore potential links between the observed anisotropy and physical properties (stellar mass, SFR, and central SMBH mass) of our galaxies, we divided the galaxies into two subsamples for each parameter: one with high and one with low values. These subsamples correspond to the top and bottom third of each parameter's distribution. The azimuthal profiles for each subsample are shown in Fig.\ \ref{fig:sig}. By visual inspection, the only profile showing a deviation from a constant behavior is that of the high-SFR subsample. To quantify this, we fit each azimuthal profile first with a constant, and then added a sinusoidal component. Then we evaluated the statistical improvement using an F-test with a 5\% acceptance threshold. This analysis confirms that the only subsample where the addition of a sinusoidal component significantly improves the fit and satisfies the F-test ($F=1.16\%$) is the one with high-SFR subsample. 

\subsection{Spectral properties}

Although the detected signal does not allow us to carry out detailed spectral analysis of the CGM emission, we assessed the basic characteristics of the emission by computing hardness ratios. To enhance the signal-to-noise ratios, we compared three regions: one along the minor axis direction ($80^\circ<\theta<100^\circ$) and two along the major axis ($0\le\theta\le80^\circ$ and $100^\circ\le\theta\le180$). We defined the hardness ratio as the ratio of counts in the soft (0.3--1~keV) and hard (1--2~keV) band: HR$=S/H$. 

We obtained HR$=1.1_{-0.7}^{+0.6}$ along the minor axis and HR$=1.4_{-1.1}^{+0.8}$ and HR$=1.5_{-1.5}^{+0.7}$ in the two major axis sectors. We note that the uncertainties were computed using the Bayesian-based tool \texttt{BEHR} \citep{behr06}. While these values are statistically consistent within uncertainties, the slightly lower HR along the minor axis suggests a harder emission, which may be caused by hotter gas temperatures. Specifically, HR$=1.1$ corresponds to a plasma temperature of $kT\approx0.65$~keV, while HR$=1.4-1.5$ suggests $kT\approx0.4$~keV. However, given the limited signal-to-noise ratios, the analysis remains inconclusive at this stage. 

\section{Comparison with \revI{TNG50}}\label{sec:sim}

\begin{figure*}[!t]
\centering
\includegraphics[width=0.32 \textwidth]{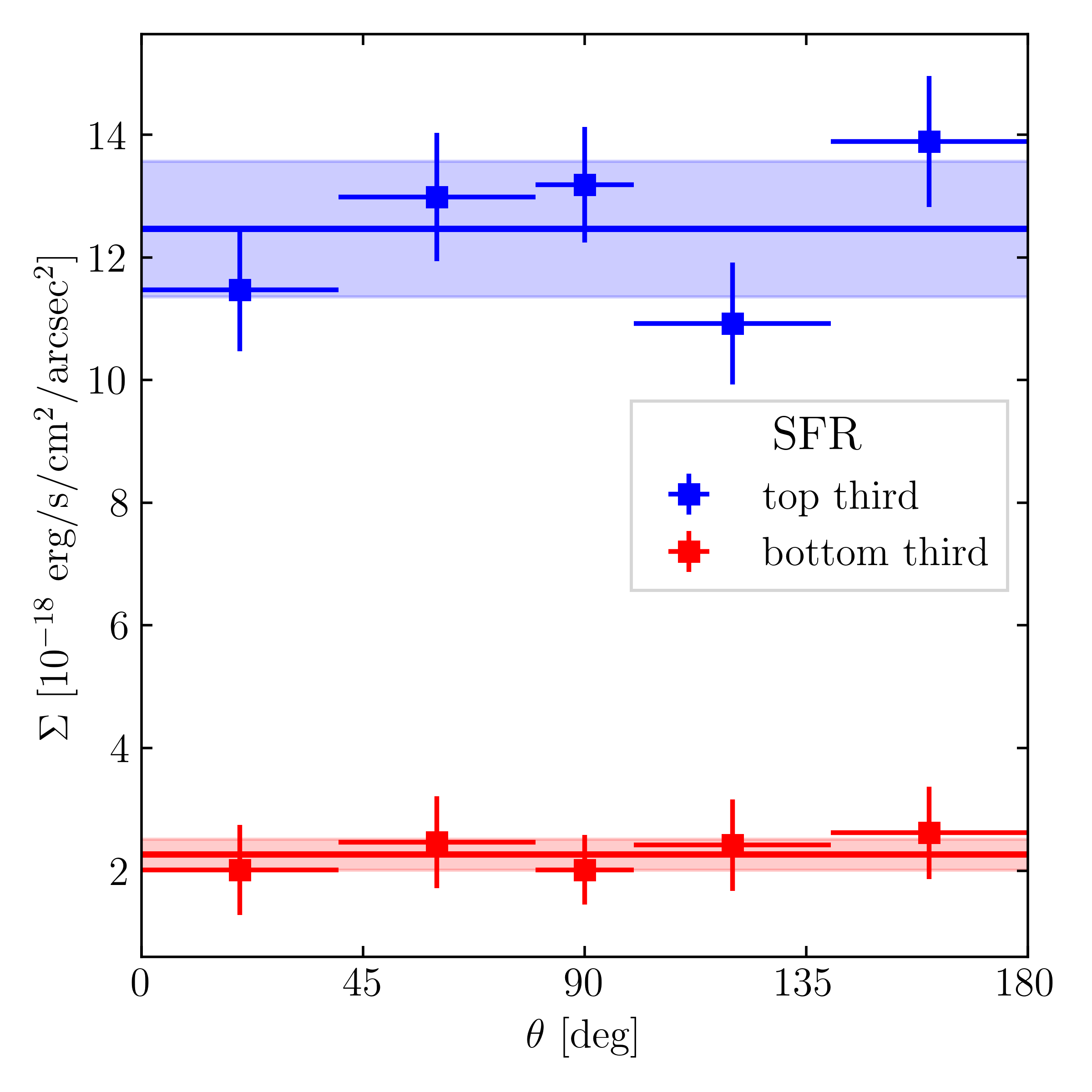}
\includegraphics[width=0.32 \textwidth]{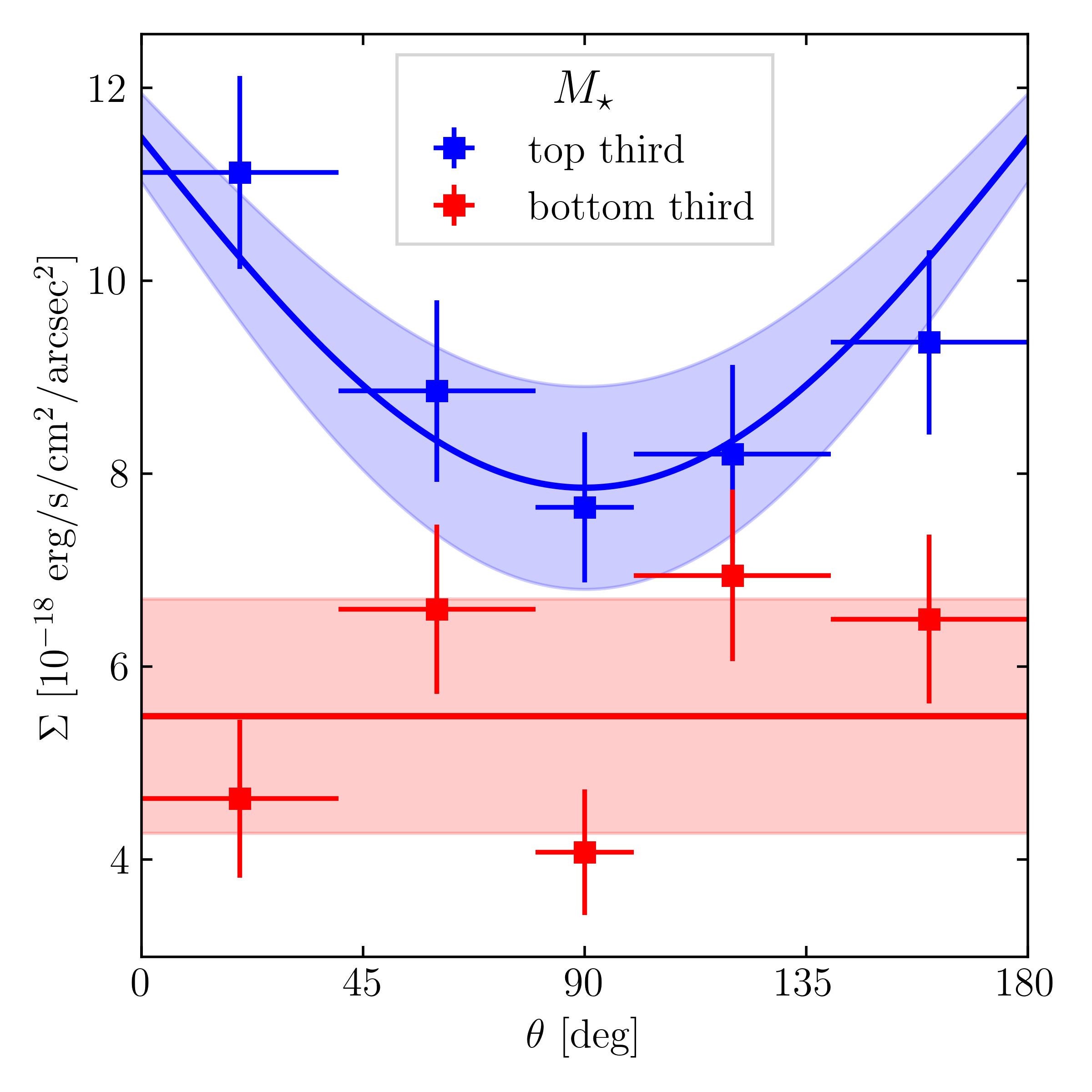}
\includegraphics[width=0.32 \textwidth]{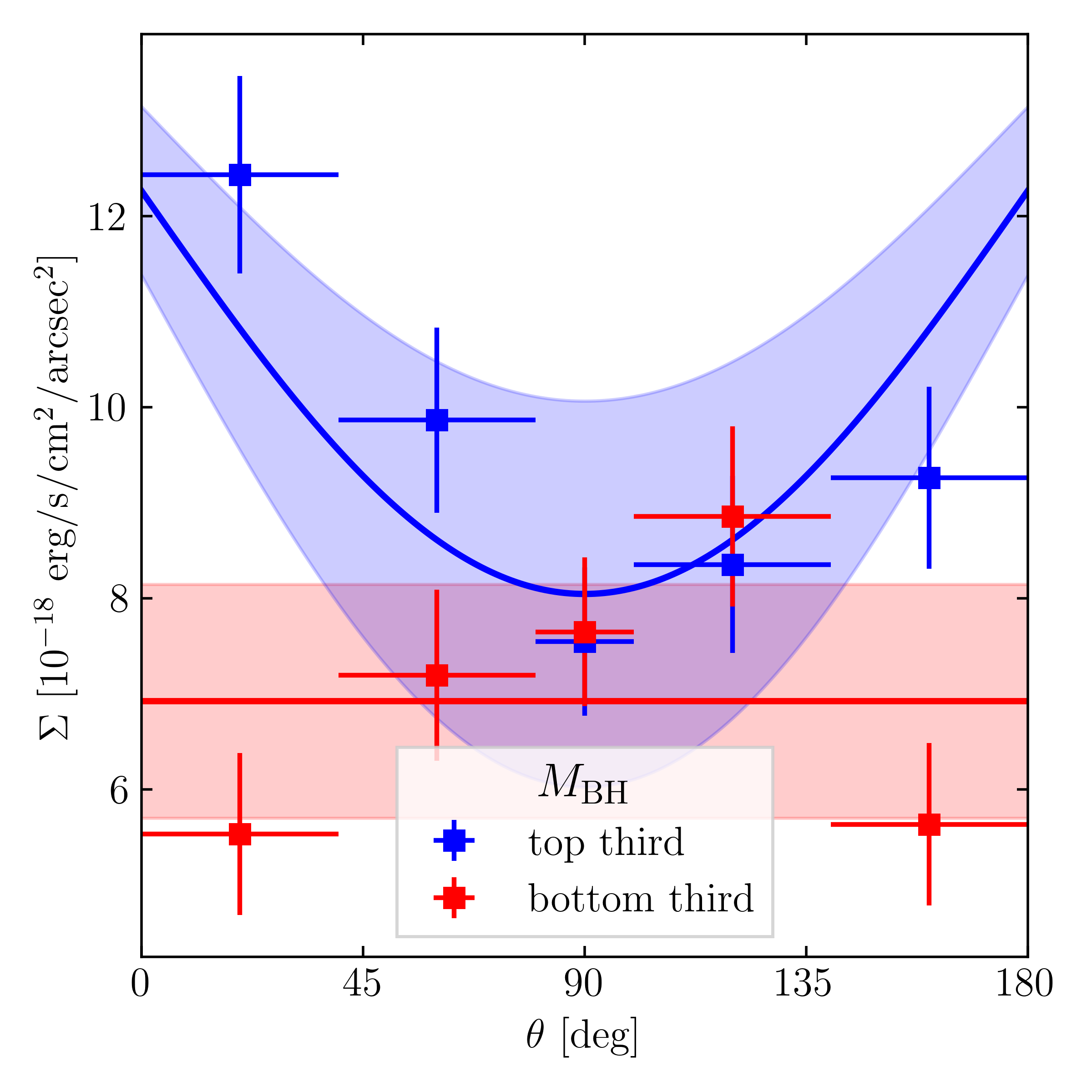}
\includegraphics[width=0.7 \textwidth]{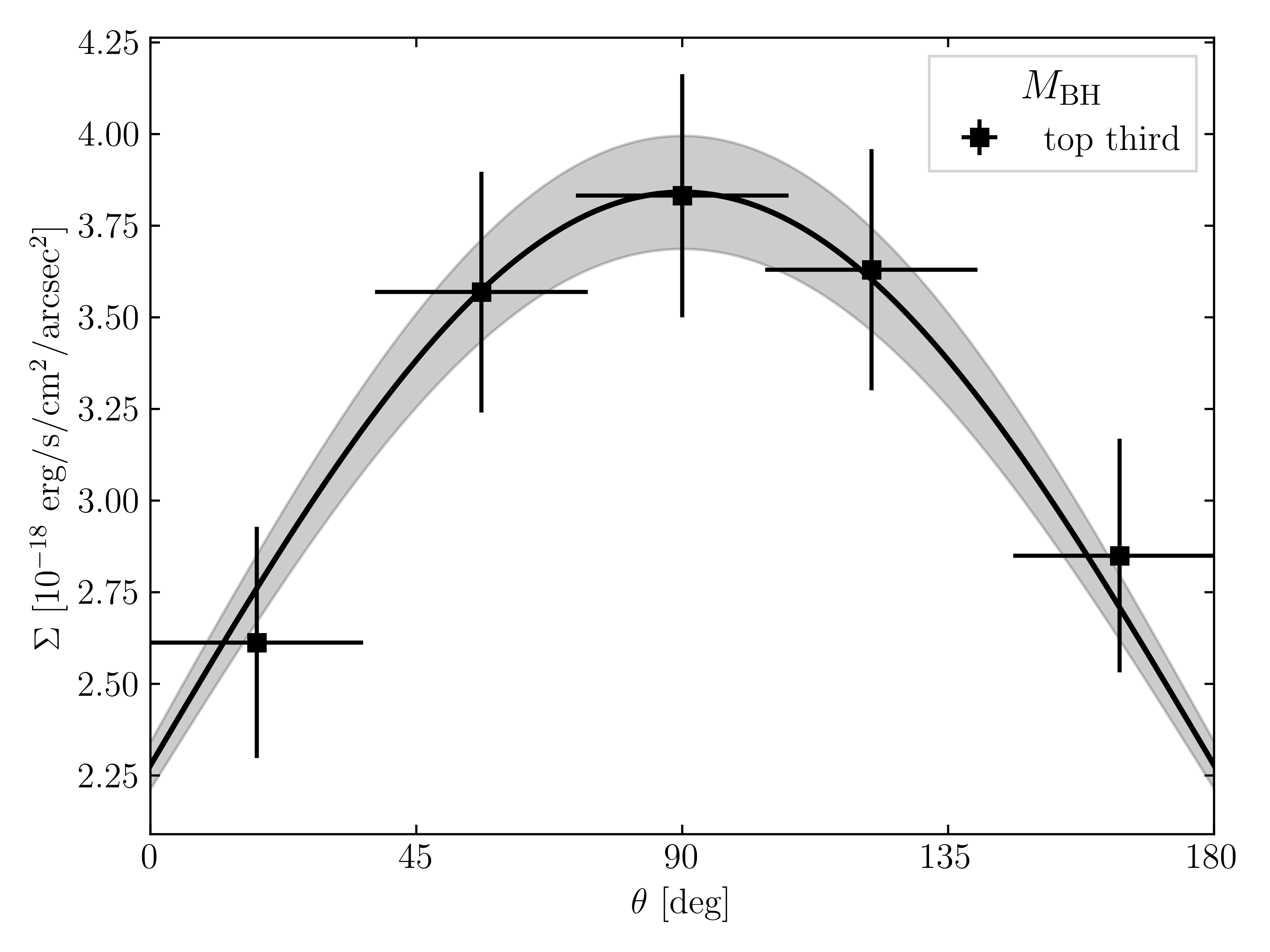}
\caption{Azimuthal profiles of different subsamples of simulated galaxies. In the upper row, the top and bottom thirds of the sample are shown in blue and red, respectively, divided based on SFR (left panel), stellar mass (middle panel), and SMBH mass (right panel). The lower panel shows the azimuthal profile of the high-SMBH mass subsample,  extracted over a larger scale. Solid lines represent the best fit, while the shaded regions denote the $1\sigma$ uncertainties.}
\label{fig:mock_sig}
\end{figure*}

In this Section, we compare these observational results with theoretical predictions derived from cosmological simulations. Specifically, we utilize a sample of simulated galaxies obtained from the TNG50 cosmological magnetohydrodynamic simulation \citep{nelson.etal.2019, pillepich.etal.2019}. The galaxies are selected at redshift ${\rm z=0}$ with their stellar masses within the observed mass range of ${\rm 10^{10.5}<M_*/M_\odot<10^{11.5}}$, resulting in a sample of 196 galaxies.   

To facilitate a proper comparison with observations, we generate \textit{Chandra} mock X-ray observations of the simulated galaxies. Starting from the simulation outputs, the CGM X-ray emission is computed based on gas cells' density, temperature, and elemental abundances. X-ray mock photons are generated by using the PyXSIM package \citep{zuhone.etal.2014}, assuming diffuse gas is an optically-thin plasma in collisional equilibrium (CIE), modeled following the approach of \cite{khabibullin.churazov.2019}. To account for galactic absorption, we apply the \texttt{TBabs} model with equivalent hydrogen column ${N_{\rm H}=2\times10^{20} \ \rm{cm^{-2}}}$. In the next step, the interaction between mock X-ray photons and the Chandra ACIS-S detector (Cycle 0) is simulated using the SOXS package \citep{zuhone.etal.2023}. For these mock observations, the simulated galaxies are virtually placed at redshift ${\rm z=0.03}$ corresponding to the median redshift of the observed galaxies. The exposure time of the simulated galaxies was set to $24$~ks, matching the mean observing time of the real galaxies. The final output of this process consists of event files, which record mock photon data in a format analogous to real observational event files. It is worth noting that for this mock exercise, we do not simulate the X-ray background and foreground. Instead, these components are added later using measurements from the observed sample. Using the actual measured background and MW foreground provides a more accurate way to incorporate these components into our mock analyses, as the foreground varies across the sky.

From the initial mass-selected sample of 196 galaxies, we further select 93 galaxies based on their SFR distribution. The sample of simulated galaxies was chosen so that the distribution of stellar mass and SFR would mimic those of our real galaxies as much as possible. The properties of the sample of simulated galaxies are reported in Fig.\ \ref{fig:mock_prop}, which shows the SFR as a function of the stellar mass, with color-coded SMBH mass. It is to be noted that the average value of SMBH mass ($\langle M_{\rm BH}\rangle\approx2.3\times10^8$~M$_\odot$) is about an order of magnitude larger than that of the sample of observed galaxies ($\langle M_{\rm BH}\rangle\approx2.0\times10^7$~M$_\odot$). This discrepancy may result from a combination of SMBH seeding and feedback processes, both stellar- and SMBH-driven, which influence the SMBH growth in TNG50 simulations \citep{truong21b}. It is worth noting that the SMBH seeding, growth, and feedback processes in TNG50 are calibrated to reproduce realistic galaxy properties such as the galaxy luminosity functions at $z=0$ \citep{pillepich18}. Therefore, rather than focusing on a single parameter like SMBH mass, it is more meaningful to assess the comparison between TNG50 simulations and observations in terms of the overall impact of SMBH feedback on the hot gaseous halos.

We analyzed the mock X-ray images following the same procedure we adopted for our sample of real galaxies. First, we compared the overall brightness of the CGM in the same circular region we adopted for the sample of observed galaxies; second, we built the azimuthal profiles. For the simulated galaxies, in the circular region of 14~kpc radius, we collected 1875 net counts corresponding to a luminosity of $4.3\times10^{39}$~erg/s. The number of counts is significantly higher than those collected for the sample of observed galaxies due to the different observation cycle of the simulated galaxies: the exposure-weighted average cycle for the observed galaxies is Cycle 9, while the mock images were produced simulating Cycle 0 instrument responses. The difference in the number of collected counts can be explained by accounting for the loss of effective area in the soft X-ray band of \cxo\ detectors due to molecular contamination of the optical blocking filter of the ACIS instrument \citep{plucinsky18,plucinsky22}. The counts were converted into luminosity using \texttt{WebPIMMS} and the same spectrum employed to create the mock X-ray images, as described above. The brightness of the CGM in the sample of mock images is in perfect agreement with that of the observed galaxies, whose luminosity amounts to $(4.2\pm0.7)\times10^{39}$~erg/s. 

We then proceeded to split our sample of simulated galaxies in the same fashion of the observed ones and compare the azimuthal profiles of the top and bottom thirds of stellar mass, SFR, and SMBH mass. These azimuthal profiles are shown in Fig.\ \ref{fig:mock_sig}. The first thing we notice is that simulated galaxies with high SFR show brighter CGM emission. This is a well-known prediction for both TNG50 and EAGLE \citep{oppenheimer20, truong.etal.2020}. This is however not supported by available X-ray data. \cxo\ observations (e.g. \citealt{bogdan11a}) and more recently eROSITA images suggest otherwise: \citet{chadayammuri22}, exploiting eFEDS data, reports that star-forming (SF) galaxies show only marginally brighter CGM emission with respect to quiescent galaxies; while \citet{zhang24c}, employing the full eROSITA dataset, does not detect any brighter CGM in SF galaxies. 

Next we inspect the azimuthal distribution for each subsample. No azimuthal profile shows enhancement in the region along the minor axis ($\theta=90^\circ$) that would resemble the observed profiles. On the contrary, the high-SMBH and stellar mass subsamples, as these two parameters are heavily correlated, show enhancement in the regions closer to the major axis  ($\theta=0^\circ,180^\circ$), and the azimuthal profiles of these subsamples can be fit with an ``inverted'' sinusoidal profile, with the minimum corresponding the region along the minor axis.

At first, this result may seem surprising, given that we know that the simulated galaxies in these subsamples host X-ray bubbles. In TNG50, the presence of MW-like bubbles is driven by SMBH feedback in the kinetic mode, which corresponds to the situation where the SMBH is accreting at relatively low rates compared to the Eddington limit (see Fig.\ 11 of \citealt{pillepich21} and \citealt{truong21}). By model construction, most massive SMBHs ($M_{\rm BH}\gtrsim10^{8.1-8.2}~{\rm M}_\odot$) are in the kinetic mode \citep{pillepich21}, hence the presence of MW-like bubbles correlates with the SMBH mass. Along the direction of the minor axis, the gas is pushed outwards creating cavities in the regions above and below the galactic center and at the same time shell-like features at their sides. These bubbles extend on scales much larger than those of the MW bubbles. Therefore, when extracting the azimuthal profiles from 14~kpc-radius regions, we do not capture the full emission of the bubbles but instead sample their lateral rims, which results in the enhancement observed in the regions closer to the galactic plane. To test this, we created an additional azimuthal profile for the high-SMBH mass subsample, extracted from a larger region extending to 50~kpc to encompass the bubbles of the simulated galaxies. This azimuthal profile, shown in the second row of Fig.\ \ref{fig:mock_sig}, as expected, shows an enhancement in the region along the minor axis.

The azimuthal profile of the high-SFR subsample does not show these features. This is because a third of the galaxies in this subsample host a SMBH accreting in thermal mode, which does not provide the feedback necessary to inflate bubble-like structures. At the same time, as mentioned above, these SF galaxies exhibit brighter CGM emission, which dilutes the signal of the galaxies hosting bubbles. These combined effects generate the observed flat azimuthal profile of these simulated galaxies.

\section{Discussion}\label{sec:disc}

The comparison with simulated galaxies provided key insights into the possible origins of the anisotropies we detected in the CGM emission of the sample of observed galaxies. The overall brightness of the detected and simulated CGM emission is compatible, but its morphology is strikingly different. As highlighted in Fig.\ \ref{fig:mock_sig}, at the scales accessible for our observed galaxies, X-ray bubbles the likes are produced by SMBH feedback in TNG50 would appear as an enhancement in the regions closer to the galactic plane rather than a signal excess in the direction along the minor axis. To obtain an enhancement in the direction of the minor axis, it is necessary to probe scales compatible with the dimensions of the bubbles themselves. This suggests that if the anisotropies we observe are caused by MW-like bubbles, these extend on scales much smaller than those reproduced by TNG50.

Another striking difference between the anisotropies we measured in observed versus simulated galaxies, is the correlation with other physical parameters. In TNG50, the presence of X-ray bubbles strongly correlates with SMBH mass. In the sample of observed galaxies, the anisotropies are correlated with the higher values of SFR. This suggests that these anisotropies, rather than SMBH feedback, are more likely related to SF phenomena.

Indeed SF-related phenomena, namely supernovae (SNe)-driven outflows cause anisotropies observed in M31 \citep{bogdan08} and M104 \citep{li11}, and starburst activity too can launch powerful outflows and winds, as observed in M82 \citep{lehnert99} and NGC~253 \citep{strickland00}. In fact, a handful of galaxies in our sample are classified as starburst galaxies, and they all fall in the top third when the sample is divided based on SFR. 

\revI{To test whether the observed properties of the gas are compatible with this scenario, adapting the results from \citet{mannucci05} following \citet{scannapieco05}, we can compute the rate of SNe (type Ia and core-collapse) based on the average values of stellar mass and star formation rate of our sample to be about 3.3 SNe per century. Assuming that each SN releases about $10^{51}$~erg, the energy input from SNe amounts to $E_\textup{SN}=7.6\times10^{41}$~erg/s. This calculation highlights that the energy of the detected CGM, characterized by an X-ray luminosity of $\approx4\times10^{39}$~erg/s, is compatible with being supplied a SNe-driven-outflows.}

In conclusion, further observational efforts are needed to clarify the origin and nature of the observed anisotropies in the CGM emission. As highlighted, any stacking procedure would average the signal from MW-like bubbles, making the unequivocal identification challenging.  To achieve a comprehensive understanding of this phenomenon, there are two viable approaches. The first is to carry out prolonged observations of a single galaxy (ideally one known to host bright CGM) with currently available X-ray observatories equipped with CCD technology and a sufficiently wide field of view (e.g.\ {\it XRISM}/Xtend). However, due to the limited spectral resolution of CCD detections, disentangling different temperature components and the kinematic signatures of the CGM will remain difficult. Therefore, the real breakthrough will come with the introduction of the next generation of X-ray telescopes, which combine high-spectral resolution, large field of view, and large effective area in the soft band. Next-generation microcalorimeter instruments would enable the tracing the CGM on significantly larger scales, the detection of bulk motions and the mapping of inflow and outflow velocities in the CGM, providing insights into the role of supernova and AGN feedback in driving bubbles or anisotropies \citep{kraft22,schellenberger24,zuhone24}. Moreover, microcalorimeters will offer precise temperature and metal abundance measurements of the large-scale CGM around MW-mass galaxies.

\section{Conclusion}\label{sec:conc}
In this paper, we studied the properties of the CGM surrounding a large sample of nearby MW-like galaxies. To achieve this, we stacked the X-ray photons from 120 \cxo\ observations of 93 individual galaxies, totaling $\sim2.2$~Ms exposure time. Our conclusions are as follows.
\begin{enumerate}
    \item We detected soft, diffuse, large-scale X-ray emission extending up to a galactocentric radius of $\sim14$~kpc, with a $0.3-2$~keV band luminosity of $4.2\times10^{39}$~erg/s.
    \item The detected emission exhibits strong anisotropies with a statistically significant enhancement along the galactic minor axis.
    \item We probed possible correlations between the CGM anisotropies and various galactic properties, including the central SMBH mass, stellar mass, and SFR. We found that the anisotropies in CGM emission correlate with higher SFR values.
    \item We explored whether these anisotropies may arise from MW-like bubbles. To test this, we compared our observational findings with mock X-ray images of galaxies from the TNG50 simulation. We found that the bubbles in the simulated galaxies extend on scales larger by a factor of $2–3$ compared to the observed features. 
    \item We speculate on the origin of the observed anisotropies. The detected features are likely associated with SF-related phenomena. Indeed, the energy input from supernovae exploding in the the bulges of the galaxies is sufficient to drive gas outflows. 
    \end{enumerate}
These results point to the necessity of further observational efforts to better explore the morphology, exact nature, and origin of the detected anisotropies. They also highlight the necessity of improving simulations to accurately reproduce the observed features and their dependence on the galactic physical parameters. 

\section*{Acknowledgments}
We thank the anonymous referee for insightful comments that improved the clarity of this work.
\'A.B. acknowledges support from the Smithsonian Institution and the Chandra Project through NASA contract NAS8-03060. The material is based upon work supported by NASA under award number 80GSFC24M0006. This research has made use of data obtained from the Chandra Data Archive and the Chandra Source Catalog, both provided by the Chandra X-ray Center (CXC). This paper employs a list of Chandra data sets, obtained by the Chandra X-ray Observatory, available at~\dataset[DOI: 10.25574/cdc.363]{https://doi.org/10.25574/cdc.363}.

\bibliography{biblio}{}
\bibliographystyle{aasjournal}

\appendix
\section{Properties of the selected sources}
We report in Table~\ref{tab:prop} the properties of the 93 galaxies composing our sample.

\begin{table*}[b]
\begin{center}   
\caption{Properties of the 93 galaxies composing our sample.}\label{tab:prop}
\begin{tabular}{lcccccccc}
\hline\hline
 ID & R.A. & DEC & $D$ [Mpc] & $\log(M_\star/{\rm M_\odot})$ & $\log$(SFR/${\rm M_\odot y}^{-1}$) & $\log(M_\textup{BH}/{\rm M_\odot})$ & ObsID & $T_\textup{exp}$  [ks] \\
 \hline
ESO018-013 & 130.9526 & -78.9494 & 76.53 & 10.93 & 0.51 & 7.1 & 4491 & 10.0 \\
IC0843 & 195.3902 & 29.1305 & 115.36 & 11.26 & 0.41 & 8.49 & 15065 & 14.9 \\
IC2475 & 141.9764 & 29.7918 & 109.26 & 10.88 & 0.14 & 7.56 & 10721 & 26.7 \\
IC2810 & 171.4377 & 14.6766 & 143.79 & 11.14 & 1.76$^{\rm SB}$ & 7.31 & 15053 & 14.8 \\
IC2810B & 171.4565 & 14.6686 & 143.18 & 10.89 & 1.16$^{\rm SB}$ & 7.32 & 15053 & 14.8 \\
IC3704 & 190.9399 & 10.77 & 129.48 & 11.02 & 1.12 & 6.67 & 8097 & 5.1 \\
NGC0120 & 6.8753 & -1.5135 & 56.23 & 10.79 & -0.38 & 7.51 & 2100 & 4.7 \\
NGC0311 & 14.3859 & 30.2808 & 70.47 & 11.08 & -0.01 & 8.68 & 3309 & 4.9 \\
\multirow{2}{*}{NGC0543} & \multirow{2}{*}{21.4582} & \multirow{2}{*}{-1.2928} & \multirow{2}{*}{99.54} & \multirow{2}{*}{10.96} & \multirow{2}{*}{-0.25} & \multirow{2}{*}{8.21} & 9583 & 9.0 \\
 &  &  &  &  &  &  & 7823 & 64.8 \\
NGC0582 & 22.9919 & 33.4765 & 61.74 & 10.9 & 0.53 & 7.34 & 9103 & 10.1 \\
\multirow{2}{*}{NGC0839} & \multirow{2}{*}{32.4284} & \multirow{2}{*}{-10.184} & \multirow{2}{*}{53.38} & \multirow{2}{*}{10.75} & \multirow{2}{*}{1.18$^{\rm SB}$} & \multirow{2}{*}{7.19} & 15666 & 29.7 \\
 &  &  &  &  &  &  & 15667 & 58.3 \\
NGC1125 & 42.9185 & -16.6507 & 43.09 & 10.62 & 0.53 & 7.28 & 14037 & 5.0 \\
NGC1204 & 46.1664 & -12.3413 & 62.24 & 10.84 & 1.12$^{\rm SB}$ & 7.28 & 10256 & 4.9 \\
NGC2315 & 105.6378 & 50.5906 & 92.4 & 11.3 & 0.47 & 7.76 & 12564 & 9.9 \\
NGC2827 & 139.8292 & 33.8808 & 102.41 & 10.78 & 0.24 & 6.84 & 5904 & 3.0 \\
NGC3341 & 160.6312 & 5.0441 & 116.63 & 11.05 & 0.98 & 7.5 & 13871 & 49.3 \\
NGC4072 & 181.0576 & 20.2097 & 96.57 & 10.71 & 0.15 & 7.45 & 12990 & 5.0 \\
NGC4180 & 183.2628 & 7.0389 & 38.25 & 10.75 & 0.58$^{\rm SB}$ & 7.19 & 9438 & 2.1 \\
NGC5675 & 218.166 & 36.3022 & 59.53 & 11.1 & 0.04 & 7.63 & 9134 & 8.9 \\
NGC5746 & 221.2331 & 1.9549 & 29.14 & 11.41 & -0.13 & 7.86 & 3929 & 36.8 \\
NGC5793 & 224.8533 & -16.6935 & 49.98 & 10.88 & 0.79 & 7.32 & 2929 & 8.6 \\
PGC004532 & 18.8369 & 0.2593 & 178.34 & 10.91 & 0.22 & 7.54 & 3204 & 37.6 \\
\multirow{2}{*}{PGC006456} & \multirow{2}{*}{26.416} & \multirow{2}{*}{-4.6807} & \multirow{2}{*}{71.04} & \multirow{2}{*}{10.55} & \multirow{2}{*}{-0.7} & \multirow{2}{*}{6.98} & 3306 & 27.9 \\
 &  &  &  &  &  &  & 3307 & 29.7 \\
PGC023051 & 123.3756 & 54.3123 & 169.41 & 11.03 & 0.74 & 7.43 & 10313 & 30.6 \\
PGC025470 & 136.1326 & 13.1612 & 117.63 & 10.56 & 0.79 & 6.26 & 13336 & 1.5 \\
PGC032934 & 164.3302 & 44.0175 & 145.36 & 10.53 & 0.52 & 6.48 & 13355 & 1.5 \\
PGC033237 & 165.3192 & 3.5629 & 163.58 & 10.73 & 0.75 & 6.91 & 7126 & 4.1 \\
PGC035831 & 173.8724 & 49.107 & 160.5 & 11.16 & 0.12 & 7.88 & 6946 & 47.3 \\
PGC035845 & 173.9332 & 49.0376 & 127.35 & 10.78 & 0.45 & 7.37 & 6946 & 47.3 \\
PGC038305 & 181.4072 & 1.8282 & 90.51 & 10.78 & -0.41 & 7.42 & 5700 & 2.1 \\
PGC047555 & 202.8824 & -2.0187 & 149.06 & 10.55 & 0.63 & 6.29 & 11879 & 19.8 \\
PGC047841 & 203.6743 & 50.4641 & 125.33 & 10.61 & -0.14 & 7.25 & 5772 & 19.5 \\
PGC048337 & 205.0114 & 40.4209 & 105.86 & 10.76 & 0.03 & 7.01 & 3223 & 47.0 \\
PGC052888 & 222.1809 & 18.2893 & 167.92 & 10.96 & 0.76 & 6.34 & 13954 & 19.8 \\
PGC057042 & 241.3142 & 17.5397 & 164.24 & 10.65 & 0.33 & 6.06 & 4996 & 21.8 \\
PGC067153 & 325.0871 & 12.3548 & 83.62 & 10.59 & 0.32 & 6.81 & 11701 & 11.4 \\
\hline\hline
\end{tabular}
\end{center}
\end{table*}

\begin{table*}
\begin{center}
\caption{Continued.}
\begin{tabular}{lccccccccc}
\hline\hline
 ID & R.A. & DEC & $D$ [Mpc] & $\log(M_\star/{\rm M_\odot})$ & $\log$(SFR/${\rm M_\odot y}^{-1}$) & $\log(M_\textup{BH}/{\rm M_\odot})$ & ObsID & $T_\textup{exp}$  [ks] \\
 \hline
PGC083454 & 167.3881 & 28.6277 & 158.92 & 11.05 & 0.8 & 7.49 & 6937 & 27.3 \\
PGC084294 & 222.0243 & 47.7885 & 157.65 & 10.54 & 0.65 & 6.1 & 4155 & 6.9 \\
PGC086659 & 237.878 & 20.2118 & 154.11 & 10.91 & 0.3 & 7.36 & 3214 & 14.9 \\
\multirow{2}{*}{PGC090491} & \multirow{2}{*}{10.2922} & \multirow{2}{*}{25.5856} & \multirow{2}{*}{131.13} & \multirow{2}{*}{10.5} & \multirow{2}{*}{0.36} & \multirow{2}{*}{6.92} & 7608 & 3.0 \\
 &  &  &  &  &  &  & 9098 & 5.0 \\
PGC091341 & 213.0421 & 6.5273 & 151.24 & 10.63 & 0.55 & 6.27 & 12169 & 4.0 \\
PGC091366 & 218.0625 & -1.1729 & 151.65 & 10.68 & 0.14 & 6.97 & 907 & 21.4 \\
PGC091445 & 231.7144 & 20.6226 & 171.58 & 10.76 & 0.47 & 6.23 & 15289 & 3.9 \\
PGC1013346 & 38.3827 & -7.7676 & 114.42 & 10.56 & 0.28 & 6.34 & 15017 & 10.0 \\
\multirow{2}{*}{PGC1156466} & \multirow{2}{*}{222.7538} & \multirow{2}{*}{0.0761} & \multirow{2}{*}{181.32} & \multirow{2}{*}{10.6} & \multirow{2}{*}{-0.66} & \multirow{2}{*}{7.15} & 14008 & 25.7 \\
 &  &  &  &  &  & & 4093 & 2.1 \\
PGC1172613 & 20.9171 & 0.6823 & 139.28 & 10.74 & 0.55 & 6.92 & 6802 & 9.8 \\
PGC1184553 & 220.7824 & 1.1 & 143.38 & 10.57 & 0.61 & 6.49 & 3960 & 11.0 \\
PGC1227743 & 165.6039 & 2.4619 & 165.89 & 10.71 & -0.39 & 6.93 & 6697 & 1.5 \\
\multirow{2}{*}{PGC1335813} & \multirow{2}{*}{201.4785} & \multirow{2}{*}{7.9757} & \multirow{2}{*}{148.35} & \multirow{2}{*}{10.57} & \multirow{2}{*}{-0.59} & \multirow{2}{*}{6.95} & 14212 & 5.0 \\
 &  &  &  &  &  & & 14214 & 4.9 \\
PGC139305 & 155.5809 & 13.0562 & 141.41 & 10.76 & 0.48 & 6.5 & 4107 & 10.0 \\
\multirow{2}{*}{PGC1408577} & \multirow{2}{*}{322.4297} & \multirow{2}{*}{12.3844} & \multirow{2}{*}{186.52} & \multirow{2}{*}{10.85} & \multirow{2}{*}{1.04$^{\rm SB}$} & \multirow{2}{*}{7.01} & 11029 & 34.2 \\
 &  &  &  &  &  &  & 11886 & 13.6 \\
PGC1429780 & 29.4271 & 13.3885 & 182.73 & 11.08 & 0.91 & 7.07 & 5631 & 7.1 \\
\multirow{2}{*}{PGC1458197} & \multirow{2}{*}{147.3594} & \multirow{2}{*}{14.4425} & \multirow{2}{*}{183.48} & \multirow{2}{*}{10.52} & \multirow{2}{*}{0.48} & \multirow{2}{*}{6.9} & 2453 & 10.6 \\
 &  &  &  &  &  & & 2095 & 13.8 \\
\multirow{2}{*}{PGC154477} & \multirow{2}{*}{149.4661} & \multirow{2}{*}{1.7554} & \multirow{2}{*}{134.35} & \multirow{2}{*}{10.85} & \multirow{2}{*}{-0.24} & \multirow{2}{*}{7.49} & 15220 & 49.9 \\
 &  &  &  &  &  & & 15244 & 47.5 \\
PGC1556460 & 222.2384 & 18.3262 & 176.54 & 10.78 & 0.11 & 7.07 & 13954 & 19.8 \\
PGC1559292 & 222.0174 & 18.4272 & 159.64 & 10.76 & -0.01 & 7.37 & 13954 & 19.8 \\
PGC1673278 & 339.0281 & 22.6236 & 160.34 & 11.06 & 0.94 & 7.51 & 9962 & 1.0 \\
PGC172710 & 9.9516 & -9.1429 & 154.47 & 10.9 & 0.8 & 7.45 & 4888 & 9.6 \\
PGC1806010 & 234.3709 & 27.3702 & 137.88 & 10.54 & 0.2 & 5.91 & 14955 & 6.0 \\
PGC1857316 & 243.449 & 29.1532 & 131.85 & 10.62 & 0.14 & 6.96 & 4799 & 4.0 \\
PGC2047202 & 219.178 & 34.3104 & 184.97 & 10.81 & -0.24 & 7.86 & 7382 & 4.8 \\
\multirow{5}{*}{PGC2048041} & \multirow{5}{*}{218.1464} & \multirow{5}{*}{34.365} & \multirow{5}{*}{178.39} & \multirow{5}{*}{10.66} & \multirow{5}{*}{-0.12} & \multirow{5}{*}{6.95} & 9896 & 50.9 \\
 &  &  &  &  &  & & 18474 & 24.6 \\
 &  &  &  &  &  & & 7384 & 4.4 \\
 &  &  &  &  &  &  & 3657 & 4.7 \\
 &  &  &  &  &  &  & 3647 & 4.6 \\
PGC2062146 & 244.1918 & 35.285 & 129.2 & 10.78 & 0.45 & 6.79 & 3340 & 5.1 \\
\multirow{6}{*}{PGC214144} & \multirow{6}{*}{206.0934} & \multirow{6}{*}{55.9513} & \multirow{6}{*}{161.65} & \multirow{6}{*}{10.75} & \multirow{6}{*}{0.33} & \multirow{6}{*}{7.03} & 809 & 44.2 \\
 &  &  &  &  &  & & 18177 & 59.9 \\
 &  &  &  &  &  & & 18178 & 30.4 \\
 &  &  &  &  &  & & 19783 & 31.8 \\
 &  &  &  &  &  & & 19784 & 33.6 \\
 &  &  &  &  &  & & 20009 & 34.3 \\
\multirow{3}{*}{PGC2148895} & \multirow{3}{*}{116.2398} & \multirow{3}{*}{39.3877} & \multirow{3}{*}{186.87} & \multirow{3}{*}{10.58} & \multirow{3}{*}{0.52} & \multirow{3}{*}{6.5} & 6111 & 49.5 \\
 &  &  &  &  &  & & 3197 & 20.2 \\
 &  &  &  &  &  & & 3585 & 19.9 \\
PGC2155980 & 254.213 & 39.8083 & 125.96 & 10.55 & 0.67 & 6.24 & 11695 & 6.1 \\
PGC2176864 & 119.3325 & 41.1752 & 171.64 & 11.1 & 0.18 & 7.44 & 3032 & 7.3 \\
PGC2208853 & 117.7662 & 42.9153 & 173.04 & 10.86 & 0.03 & 7.57 & 6814 & 4.0 \\
\hline\hline
\end{tabular}
\end{center}
\end{table*}

\begin{table*}
\begin{center}
\caption{Continued.}
\begin{tabular}{lccccccccc}
\hline\hline
 ID & R.A. & DEC & $D$ [Mpc] & $\log(M_\star/{\rm M_\odot})$ & $\log$(SFR/${\rm M_\odot y}^{-1}$) & $\log(M_\textup{BH}/{\rm M_\odot})$ & ObsID & $T_\textup{exp}$ [ks] \\
 \hline
\multirow{5}{*}{PGC2296674} & \multirow{5}{*}{198.7014} & \multirow{5}{*}{47.3516} & \multirow{5}{*}{130.85} & \multirow{5}{*}{10.73} & \multirow{5}{*}{0.47} & \multirow{5}{*}{6.77} & 14382 & 9.8 \\
 &  &  &  &  &  & & 14398 & 9.9 \\
 &  &  &  &  &  & & 14420 & 20.1 \\
 &  &  &  &  &  & & 19296 & 29.9 \\
 &  &  &  &  &  & & 20977 & 9.9 \\
PGC2463793 & 123.4321 & 54.2403 & 166.26 & 10.77 & 0.55 & 6.8 & 10313 & 30.6 \\
PGC2483087 & 212.2681 & 54.8722 & 175.9 & 10.67 & -0.49 & 7.63 & 14910 & 17.7 \\
PGC2544828 & 168.5492 & 56.6966 & 142.51 & 10.5 & 0.73 & 6.46 & 6965 & 5.0 \\
PGC2578440 & 172.4031 & 58.4148 & 175.9 & 10.6 & 0.52 & 6.93 & 6227 & 10.2 \\
PGC3089053 & 126.1804 & 29.9899 & 107.61 & 10.64 & 0.43 & 6.91 & 23708 & 9.5 \\
PGC991007 & 359.9656 & -9.4427 & 173.38 & 10.67 & 0.34 & 7.14 & 9191 & 3.9 \\
UGC00330 & 8.4241 & 39.5447 & 84.82 & 10.9 & -0.24 & 7.35 & 12991 & 34.6 \\
UGC01396 & 28.9025 & 21.3175 & 68.8 & 10.65 & 0.09 & 7.09 & 7448 & 4.9 \\
UGC01911 & 36.6525 & 0.6488 & 166.97 & 11.06 & 0.68 & 6.95 & 8619 & 9.0 \\
\multirow{2}{*}{UGC01934} & \multirow{2}{*}{36.9643} & \multirow{2}{*}{0.5015} & \multirow{2}{*}{179.47} & \multirow{2}{*}{11.23} & \multirow{2}{*}{0.64} & \multirow{2}{*}{7.69} & 8626 & 8.9 \\
 &  &  &  &  &  & & 16303 & 49.4 \\
UGC02456 & 44.9942 & 36.8205 & 51.52 & 10.81 & 1.16$^{\rm SB}$ & 6.8 & 4075 & 19.9 \\
\multirow{2}{*}{UGC02626} & \multirow{2}{*}{49.2489} & \multirow{2}{*}{41.3565} & \multirow{2}{*}{91.94} & \multirow{2}{*}{10.96} & \multirow{2}{*}{-0.21} & \multirow{2}{*}{8.07} & 5597 & 25.2 \\
 &  &  &  &  & & & 8473 & 29.7 \\
UGC02639 & 49.4601 & 41.9676 & 55.25 & 10.5 & -0.54 & 6.59 & 17278 & 4.7 \\
UGC02645 & 49.3604 & -0.0942 & 97.73 & 10.77 & 0.52 & 7.07 & 10267 & 10.9 \\
UGC03326 & 84.9046 & 77.3125 & 76.6 & 11.15 & 0.74 & 7.6 & 13810 & 79.0 \\
UGC04387 & 126.187 & 46.9072 & 166.47 & 11.2 & 0.63 & 7.31 & 15159 & 8.0 \\
UGC07196 & 182.9973 & 15.4013 & 121.72 & 11.38 & 0.06 & 7.7 & 2104 & 4.9 \\
\multirow{2}{*}{UGC08186} & \multirow{2}{*}{196.4943} & \multirow{2}{*}{3.9565} & \multirow{2}{*}{96.89} & \multirow{2}{*}{10.98} & \multirow{2}{*}{0.43} & \multirow{2}{*}{7.29} & 3358 & 8.2 \\
 &  &  &  &  &  &  & 3966 & 118.2 \\
UGC08515 & 202.9626 & 11.2507 & 101.76 & 10.8 & -0.37 & 7.27 & 3213 & 30.7 \\
UGC09224 & 215.977 & 34.7236 & 152.06 & 10.54 & 0.32 & 6.2 & 3636 & 4.7 \\
UGC09233 & 216.1458 & 35.2798 & 108.27 & 10.89 & 0.54 & 6.95 & 3620 & 4.7 \\
UGC09284 & 217.0975 & 33.2537 & 61.83 & 10.56 & -0.1 & 7.13 & 4254 & 4.5 \\
UGC10417 & 247.42 & 40.687 & 132.19 & 10.9 & 0.23 & 7.34 & 12816 & 4.9 \\
UGC12589 & 351.257 & 0.0006 & 142.86 & 10.87 & 1.2 & 7.18 & 8612 & 8.8 \\ 
\hline\hline\\
\end{tabular}
\end{center}
{\bf Notes.} Identifier, R.A., DEC, distance, stellar mass, and SFR were taken from the HECATE catalog. An apex SB to the value of SFR indicates if the galaxy is classified as starburst following \citet{strickland04b} criterion. The SMBH mass is inferred from either the central velocity dispersion or the stellar mass if the former is not available (see Sec. \ref{sec:sample} for details). The ObsID is the unique identifier of the \cxo\ observation of which we report the exposure time.
\end{table*}

\end{document}